\documentclass[fleqn,usenatbib]{mnras}

\usepackage[T1]{fontenc}
\usepackage{ae,aecompl}

\usepackage{graphicx}        
\usepackage{amsmath}        
\usepackage{amssymb}        
\usepackage{newtxtext,newtxmath}

\usepackage{subfigure}
\usepackage{multirow}
\usepackage{tabularx}
\usepackage{multicol}
\usepackage{soul}
\usepackage{xcolor}

\title[Simulations of bar formation in the Local Group]{Numerical simulations of bar formation in the Local Group}

\author[Ornela F. Marioni et al.]{
Ornela F. Marioni,$^{1,2,3}$\thanks{E-mail: ornela.marioni@unc.edu.ar}
Mario G. Abadi, $^{1,3}$
Stefan Gottl\"ober, $^{4}$
Gustavo Yepes $^{5,6}$
\\
$^{1}$Instituto de Astronom\'ia Te\'orica y Experimental, UNC-CONICET, Laprida 854, X5000BGR C\'ordoba, Argentina.\\
$^{2}$Facultad de Matem\'atica, Astronom\'ia, F\'isica y Computaci\'on, UNC, Av. Medina Allende s/n, X5000HUA C\'ordoba, Argentina.\\
$^{3}$Observatorio Astron\'omico de C\'ordoba, Universidad Nacional de C\'ordoba, Laprida 854, X5000BGR C\'ordoba, Argentina.\\
$^{4}$Leibniz-Institut f\"ur Astrophysik Potsdam (AIP), An der Sternwarte 16, D-14482 Potsdam, Germany.\\
$^{5}$ Departamento de F\'isica Te\'orica, M\'odulo 8, Facultad de Ciencias, Universidad Aut\'onoma de Madrid, E-28049 Madrid, Spain.\\
$^{6}$ CIAFF, Facultad de Ciencias, Universidad Aut\'onoma de Madrid, E-28049 Madrid, Spain. \\
}

\date{Accepted XXX. Received YYY; in original form ZZZ}

\pubyear{2022}

\begin{document}
\label{firstpage}
\pagerange{\pageref{firstpage}--\pageref{lastpage}}

\maketitle

\begin{abstract}

More than 50 per cent of present-day massive disc galaxies show a rotating stellar bar. Their formation and dynamics have been widely studied both numerically and observationally. Although numerical simulations in the $\Lambda$CDM cosmological framework predict the formation of such stellar components, there seems to be a tension between theoretical and observational results.
Simulated bars are typically larger in size and have slower pattern speed than observed ones. We study the formation and evolution of barred galaxies, using two $\Lambda$CDM zoom-in hydrodynamical simulations of the CLUES project that follow the evolution of a cosmological Local Group-like volume. We found that our simulated bars, at $z = 0$, are both shorter and faster rotators than previous ones found in other studies on cosmological simulations alleviating the tension mentioned above. These bars match the short tail-end of the observed bar length distribution. In agreement with previous numerical works, we find that bars form in those systems where the disc self-gravity is dominant over the dark matter halo, making them unstable against bar formation. Our bars developed in the last 3-4~Gyr until they achieve their current length and strength; as bars grow, their lengths increase while their rotation speeds decrease. Despite this slowdown, at redshift $z = 0$ their rotation speeds and size match well the observational data.

\end{abstract}




\section{Introduction}
Barred galaxies constitute about two-thirds of the disc galaxy population observed in the Local Universe with approximately half of them showing evidence of a strong bar \citep{Sellwood&Wilkinson_1993}. 

Bars are found to have different sizes and shapes. They vary from large structures that dominate the entire disc to small oval distortions in the inner regions of the galaxies.These structures play a significant role in galaxy evolution, particularly in the redistribution of angular momentum between the disc and the dark matter halo \citep{Debattista&Sellwood_2000, Athanassoula2003}.

Roughly speaking, two main parameters are commonly defined in order to characterise bar properties: bar length and pattern speed. Bar lengths can range from $l_{\rm bar} \sim\! 1\,{\rm kpc}$ (called short bars) up to $l_{\rm bar} \sim\!10\,{\rm kpc}$ (called long bars).

Bars are formed by instabilities in the disc. These instabilities may be produced by internal mechanisms that act over long timescales, redistributing mass in the inner regions of the galaxy and, therefore, their angular momentum. 
Or they may also be produced by external processes, like mergers and flybys that can delay or advance the bar formation \citep{Moetazedian&Just2016,Zana_etal2018}. 

\citet{Miller&Prendergast1968} and \citet{Hockney&Hohl1969} were the pioneers who ran N-body simulations of discs of collisionless particles. In their simulations, they found that it is easy to form unstable discs that form a bar over a dynamical time-scale. On the other hand, it is difficult to build perfect rotationally supported discs free from instabilities. Although the global instabilities explain well the bar structure formation, there is not yet a good understanding of stable discs.

To stabilise these fragile discs, avoiding bar formation, \citet{Ostriker&Peebles1973} proposed the addition of a massive non-rotating spherical halo surrounding the disc. Further on \citet{Efstathiou1982} implemented this idea using live dark matter haloes.

Early analytical and numerical work by \citet{Weinberg1985} showed that bars suffer dynamical friction against their surrounding haloes, which slows them down
between bar and halo has been further developed by \citet{Little&Carlberg_1991, Hernquist&Weinberg_1992, Sellwood&Debattista2006,Weinberg&Katz_2007a, Weinberg&Katz_2007b} and \citet{Sellwood2008}, among others.

Many studies of idealised models have focused on the analysis of bar formation and evolution \citep{Combes_etal1990,Debattista&Sellwood_2000,Athanassoula&Misiriotis2002}. They find that, as bars develop, they slow down and grow longer and stronger. More recently, studies with hydrodynamical cosmological simulations have been performed, such as, \citet{Scannapieco&Athanassoula2012, Okamoto_etal2016, Algorry_etal2017, Peschken&Lokas2019}. Their results are not in complete agreement with observations since most of these results find that bars are strongly braked throughout their evolution, resulting in slow bars not consistent with the observations.

A commonly used method to classify bars as fast or slow rotators is the ratio between the corotation radius and the bar length $\mathcal{R} = \mathrm{R_{corot}} / l_{\rm bar}$. The corotation radius is the radius where the angular velocity of a circular orbit equals the bar angular velocity $\rm{\Omega_{bar}}=V_c(R_{\rm{corot}})/R_{\rm{corot}}$. If $1.0<\mathcal{R}<1.4$, the bar is considered \textit{fast}, on the other hand, if $R>1.4$ then the bar is considered \textit{slow}. 
Using theoretical arguments, \citet{Contopoulos1980} shows that a bar should be always inside the corotation radius, and thus a bar with $\mathcal{R}<1.0$ will be physically impossible.
Many observational results have demonstrated that bars have $\mathcal{R} \approx 1.4$ \citep{Corsini_2011, Aguerri_etal2015, Font_etal2017}. Meanwhile, previous numerical results with cosmological simulations \citep{Algorry_etal2017, Peschken&Lokas2019} have shown that bars slow down excessively, giving values of $\mathcal{R} \approx 3-4$.
This issue has created a supposed conflict between the existence of fast bars in the $\Lambda$CDM paradigm, calling into question this concordance cosmological model.
A recent exception is the work of \citet{Fragkoudi_etal2021} who, using the Auriga simulations, has shown that fast rotating bars can form in the $\Lambda$CDM model in agreement with observational studies. 
Our zoom-in simulations allow us to study the short tail-end of observational bars in complement with this previous work.

This paper is organised as follows: in Section~\ref{sec:simulations} we present the simulations analysed in this work and the sample selected. In Section~\ref{sec:analysisresults}, we show the main properties of the bars: bar strength, bar length and pattern speed, and comment on the main results. In Section~\ref{sec:summary}, we include the summary and conclusions of our work.

\section{Simulations} 
\label{sec:simulations}

We analyse two sets of simulated galaxies from two zoom-in high-resolution simulations that are part of the CLUES-project \footnote{Constrained Local UniversE Simulations:\\ \url{https://www.clues-project.org/cms/}} \citep{Gottlober_etal2010,Yepes_etal2014}. This collaboration aims to run N-body plus hydrodynamical cosmological simulations, mimicking the observational properties of the Local Group of galaxies. Observational data is used to constrain initial conditions to resemble the final mass distribution and velocity field of the Local Group. 
These observational constraints are set up from peculiar velocities from the MARK III \citep{Willick_etal1997}, surface brightness fluctuations from \citet{Tonry_etal2001}, the catalogue of neighbouring galaxies \citep{Karachentsev_etal2004} and the catalogue of nearby X-ray selected clusters of galaxies \citep{Reiprich&Bohringer_2002}. Then, the initial conditions are generated as constrained realisations of Gaussian fields employing the \citet{Hoffman&Ribak_1991} algorithm in a $256^3$ uniform mesh.
Even though the initial conditions of the simulations are designed to reproduce the large-scale environment of the Local Group, at smaller scales ($\lesssim 1 h^{-1}\,\rm{Mpc}$) they remain mainly random. Therefore, several runs have been made to obtain the correct Local Group candidate (i.e. two haloes located at a proper distance with their correct relative velocities, masses, etc.)

These initial conditions are used to first run a cosmological dark matter-only simulation in a box of $64 h^{-1}\,\rm{Mpc}$ size with $1024^3$ particles with cosmological parameters ($\Omega_{\Lambda}=0.76$, $\Omega_{m}=0.24$, $\Omega_{b}=0.042$, $h=0.73$, $\sigma_8=0.75$ and $n=0.95$) given by WMAP3 \citep{Spergel2007}. Then, a smaller approximately spherical region of $2 h^{-1}\,\rm{Mpc}$ radius is re-simulated at a higher resolution using the \citet{Klypin_etal2001} zoom-in technique with $4096^3$ effective dark matter-particles adding the same number of gas particles. This region is selected, at redshift $z=0$, in the original box to enclose the three most massive dark matter haloes at the position of the Local Group candidate.

Both zoom-in simulations start from the same initial conditions but are evolved with two different hydrodynamical codes: TreePM+SPH {\footnotesize GADGET-2} code \citep{Springel2005} and N-body+SPH {\footnotesize GASOLINE} code \citep{Wadsley2004}.

\begin{table}
\caption[]{Final ($z=0$) simulation parameters of the high-resolution region. In column (1) we have the simulation code. Then in columns (2), (3) and (4) are the dark matter, gas and star particles masses, and in columns (5) and (6) there are the softening radius of dark matter $\epsilon_{\rm DM}$ and baryon $\epsilon_{\rm BAR}$ particles respectively.}
\label{tab:sim_parameters}
\centering
\begin{tabular}{cccccc}
\hline \noalign{}
Simulation & $\rm{m_{DM}}$ & $\rm{m_{GAS}}$ & $\rm{m_{STR}}$ & $\epsilon_{\rm DM}$ & $\epsilon_{\rm BAR}$ \\
code & [$10^{5} \rm{M_{\odot}}$] & [$10^{4} \rm{M_{\odot}}$] & [$10^{4} \rm{M_{\odot}}$] & [kpc] & [kpc]\\
\hline \noalign{}
{\footnotesize GADGET-2} & 2.87 & 6.06 & 3.02 & 0.14 & 0.14 \\
{\footnotesize GASOLINE} & 2.87 & 6.06 & 1.45 & 0.49 & 0.22 \\
\hline 
\end{tabular}
\end{table}

The CLUES-GADGET2 simulation follows radiative cooling, star formation, supernova feedback, etc. as described by \citet{Springel&Hernquist2003}.
The interstellar medium is modelled as a two-phase medium composed of hot and cold gas affected after redshift $z = 6$ by an uniform but evolving cosmic ionising ultraviolet background \citep{Haardt&Madau1996}.
Cooling is independent of the medium metallicity and there is no cooling below $10^4\,{\rm K}$. 
The gas is enriched by supernova explosions. Gas particles spawn into stars when they reach certain conditions. Each gas particle can spawn on two star particles only. When the first star is born, it has half the mass of its progenitor gas particle and this particle loses half of its mass. When the second star is born, the progenitor disappears to preserve the total mass. Star particles interact with other particles in the same way as dark matter particles do (i.e., as collisionless particles), and only a small fraction $\beta = 0.12$ will explode as supernovae.
The CLUES-GADGET2 simulation has been described in more detail by \citet{Libeskind_etal2010}. It has also been analysed in several other works, e.g. \citet{Libeskind_etal2011, Knebe_etal2011, BenitezLl_etal2013, Benitez-Llambay_etal2015}.

\begin{table*}
\caption[]{Main properties of simulated galaxies. From left to right: virial mass, stellar mass, virial radius, half mass (stellar) radius, bar length, bar pattern speed, corotation radius, quotient between the mean corotation radius and mean bar length, bar formation time.}
\label{tab:properties}
\centering
\begin{tabular}{lccccccccc}
\hline\noalign{}
        Galaxy name & ${\rm M_{vir}}$ & ${\rm M_{stellar}}$ & ${\rm r_{vir}}$ & ${\rm r_{50}}$ & $l_{\rm bar}$ & ${\rm \Omega_{bar}}$ & $R_{\rm corot}$ & $\bar{\mathcal{R}}$ & $t_{\rm bar}$ \\
        
         & [$10^{11} {\rm M_{\odot}}$] & [$10^{10} {\rm M_{\odot}}$] & [${\rm kpc}$] & [${\rm kpc}$] & [${\rm kpc}$] & [${\rm km\:s^{-1} kpc^{-1}}$] & [${\rm kpc}$] & & [${\rm Gyr}$]\\
\hline
        {A-Gadget2}  & 6.15 & 1.33 & 167.2 & 1.61 & $1.40_{-0.14}^{+0.13}$ & $71.43_{-2.57}^{+3.75}$ & $2.13_{-0.09}^{+0.06}$ & 1.52 & 8.68 \\ 
        {A-Gasoline} & 5.93 & 1.11 & 166.8 & 0.87 & $0.90_{-0.01}^{+0.02}$ & $99.16_{-6.13}^{+3.20}$ & $1.59_{-0.05}^{+0.09}$ & 1.76 & 8.09\\ 

        {B-Gadget2}  & 4.69 & 1.38 & 153.5 & 2.40 & - & - & - & - & - \\
        {B-Gasoline} & 4.59 & 1.03 & 150.5 & 2.24 & $1.45_{-0.24}^{+0.37}$ & $32.67_{-0.85}^{+0.48}$ & $3.79_{-0.06}^{+0.10}$ & 2.61 & 10.62 \\

        {C-Gadget2}  & 2.36 & 0.79 & 123.1 & 3.74 & - & - & - & - & - \\
        {C-Gasoline} & 2.28 & 0.48 & 119.6 & 2.58 & - & - & - & - & - \\
\hline 
\end{tabular}
\end{table*}

The CLUES-GASOLINE simulation uses the physics implemented in \citet{Governato2010} and \citet{Guedes2011}. It includes star formation, gas cooling, supernova feedback and a cosmic ultraviolet background of \citet{Haardt&Madau1996}.
Star formation proceeds differently from the other simulation. When the gas becomes cold and dense, star formation can take place following the Schmidt law with a star formation rate $\propto \! \rho^{1.5}$.
When these stars die, they enrich the interstellar medium with metals, and the returned gas has the same metallicity as the dead star. Energy feedback is implemented following \citet{Stinson_etal2006}.
The number of stars that explode as supernovae and their total mass are obtained from stellar lifetime calculations from \citet{Raiteri_etal1996}.
The CLUES-GASOLINE simulation has been described in more detail by \citet{Santos-Santos_etal2016}.  It has also been analysed in \citet{Santos-Santos_etal2017} and \citet{Mostoghiu_etal2018}.

In Table~\ref{tab:sim_parameters}, we list the final ($z=0$) particle masses and softening lengths of each high-resolution simulation. Note that the difference in stellar masses is due to the different physical processes of star formation that each simulation uses. For CLUES-GADGET2 simulation, all stellar particles have the same mass, half that of the gas particle. Meanwhile, in the CLUES-GASOLINE simulation, stellar particles have different masses, in this case, we quote in Table \ref{tab:sim_parameters} the median of the masses.
The softening radius is fixed in comoving coordinates in the CLUES-GASOLINE simulation, while, in the CLUES-GADGET2 simulation, it is comoving until $z=4$ and then remains fixed in physical coordinates. We have confirmed that all the distances that we are dealing with are longer than the softening radius for all redshifts.

For both simulations, haloes have been identified using the MPI+OpenMP hybrid halo finder AHF \citep{Knollmann&Knebe2009}. This implementation places overdensities in an adaptive smoothed density field as probable halo centres. For each overdensity, the potential minimum is calculated and the gravitational bound particles are determined. Only the peaks with $\gtrsim 20$ particles are considered haloes that will be studied later. This algorithm identifies hierarchical substructures automatically as haloes, subhaloes, sub-subhaloes, etc. A subhalo is defined as a halo that lives in a more massive halo.
The merger tree was built following the ten most bound particles of each halo back through the snapshots, and associating these particles to the nearest halo centre.

\begin{figure*}
        \centering
    \subfigure[]{\label{fig:face_on}\includegraphics[width=\columnwidth]{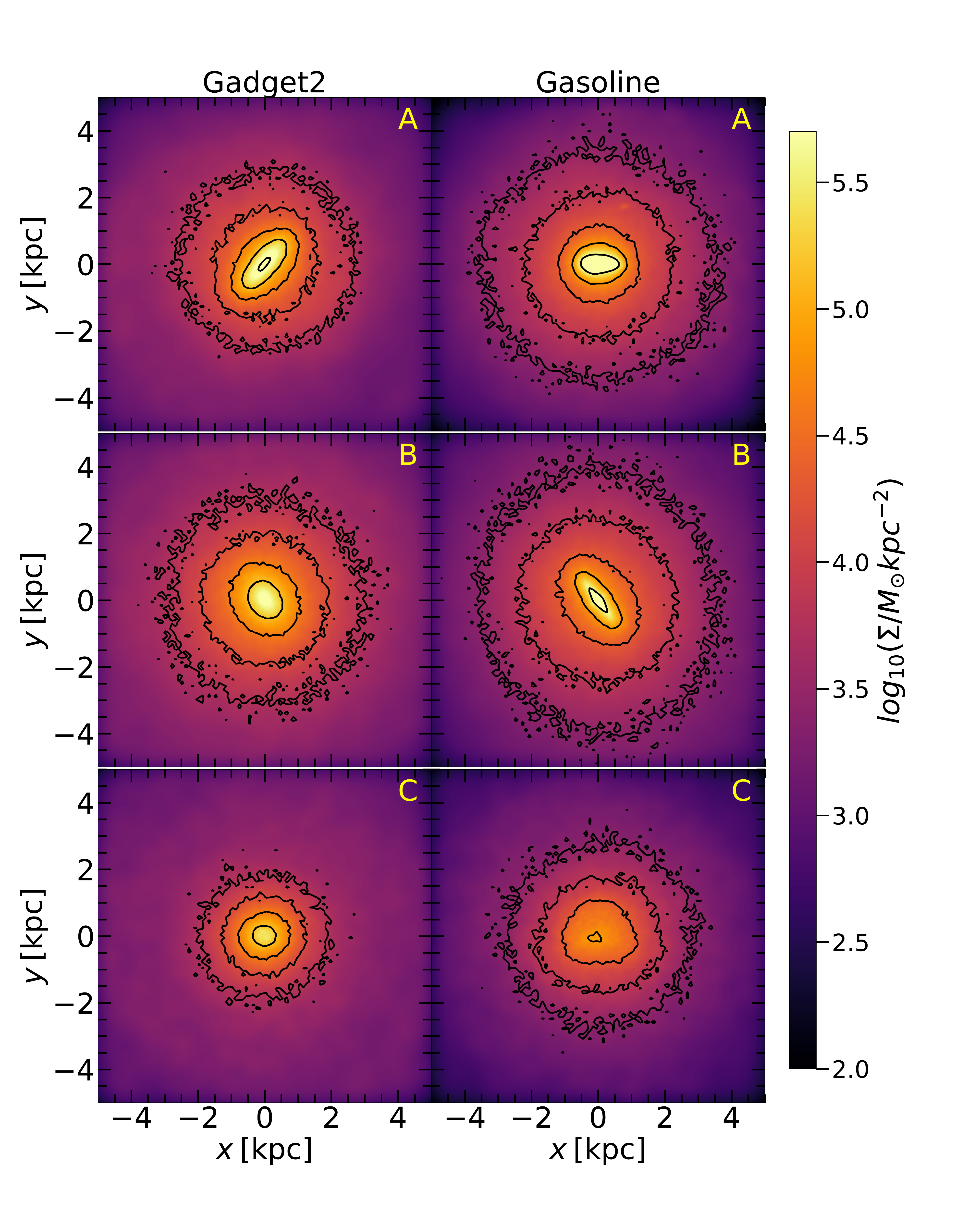}}
          \subfigure[]{\label{fig:edge_on}\includegraphics[width=\columnwidth]{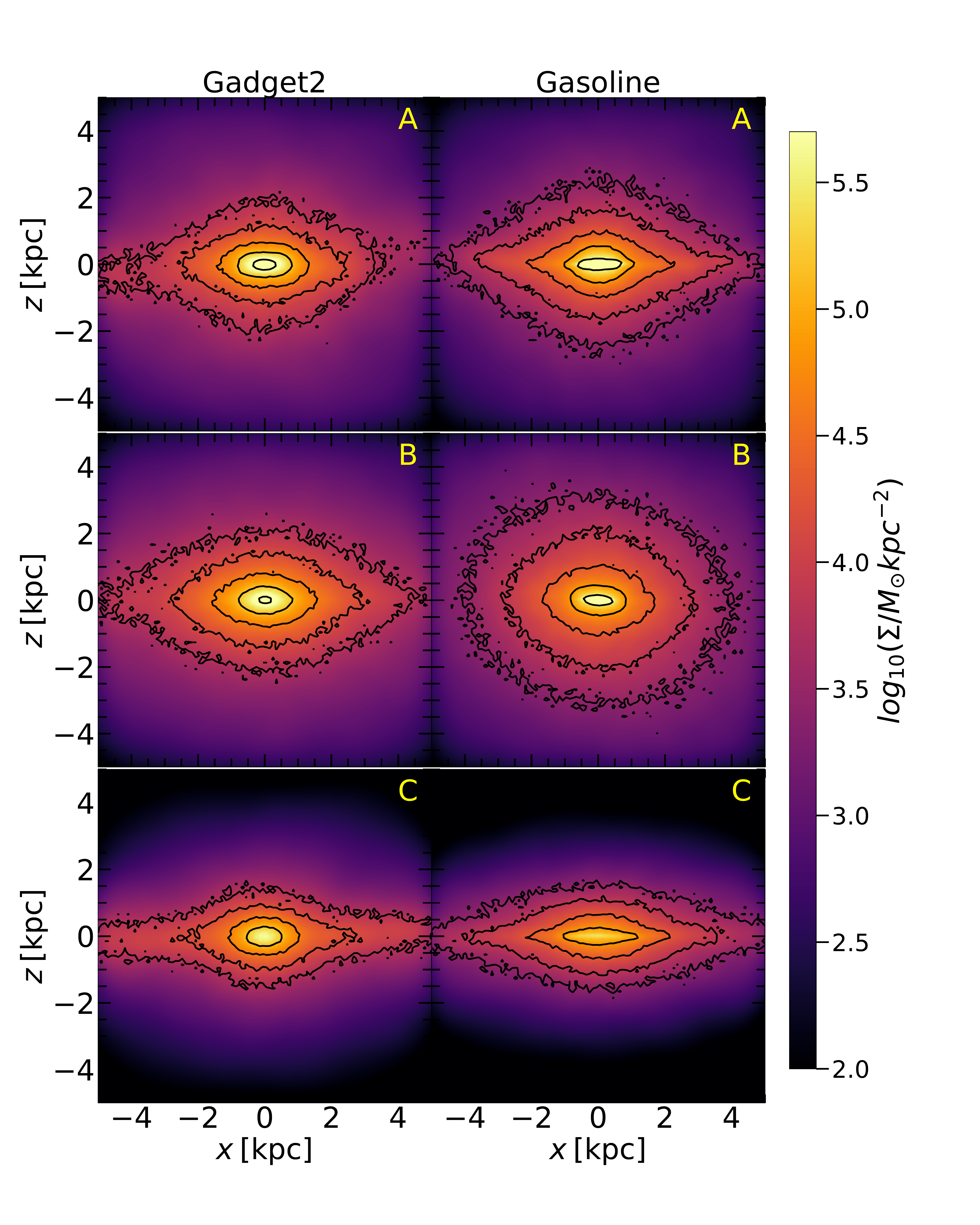}}
        \caption{Projected stellar mass density $\Sigma$: (a) face-on and (b) edge-on views at redshift $z=0$ of the central regions of the galaxies in a box of side $10\,{\rm kpc}$. On the left of each figure, there are galaxies of CLUES-GADGET2 and on the right, galaxies of CLUES-GASOLINE. The solid lines correspond to five density isocontours chosen arbitrarily at $log_{10}(\Sigma)=3.3$, $3.7$, $4.2$, $4.7$ and $5.5$.}
    \label{fig:bars}
\end{figure*}


\subsection{Galaxy sample}

At redshift $z=0$, we select the three most massive haloes (named A, B and C) identified in each simulation (CLUES-GADGET2 or CLUES-GASOLINE).
We check that these haloes are free from contaminating low-res dark matter particles.
We list the main properties of these galaxies in Table~\ref{tab:properties}.
Virial radius ${\rm r_{vir}}$ is defined as the radius where the average integrated mass density is $\Delta=200$ times the critical density of the Universe and virial mass ${\rm M_{vir}}$ is the total mass inside ${\rm r_{vir}}$. 
Following \citet[][see also \citet{BenitezLl_etal2013, Benitez-Llambay_etal2015, Ferrero_etal2017}]{Sales_etal2010} we define the galaxy radius $r_{\rm {gal}}$ as $0.15 r_{\rm{vir}}$; this is the radius that includes most of the stellar particles contained inside the dark matter halo ($\gtrsim \! 85\%$ in our galaxies) and there is no substructure inside; i.e., no subhalo is found by the AHF algorithm inside the galaxy radius.
Stellar mass ${\rm M_{stellar}}$ is computed inside a sphere of radius ${\rm r_{gal}}$. 
The half-mass radius ${\rm r_{50}}$ is the radius enclosing half of the stellar mass.

It should be noted that these three main galaxies and their corresponding dark matter haloes are systematically less massive (by a factor 2-3 in virial mass and by a factor 5-7 in stellar mass) than the current estimates of the masses of the three main spiral galaxies of our Local Group (namely Andromeda, Milky Way and Triangulum). However, they do fit \citep{BenitezLl_etal2013} the abundance matching relation \citep{Guo2010} and can be considered as normal disc-like galaxies suitable for analysing the formation of barred stellar components in this cosmological context.
We would like to point out that a comparison between simulations is beyond the scope of this paper. Each galaxy is analysed as an individual system. However, it should be noted that these galaxies cannot be used as an independent sample because they are generated using the same initial conditions.

\begin{figure}
        \includegraphics[width=\columnwidth]{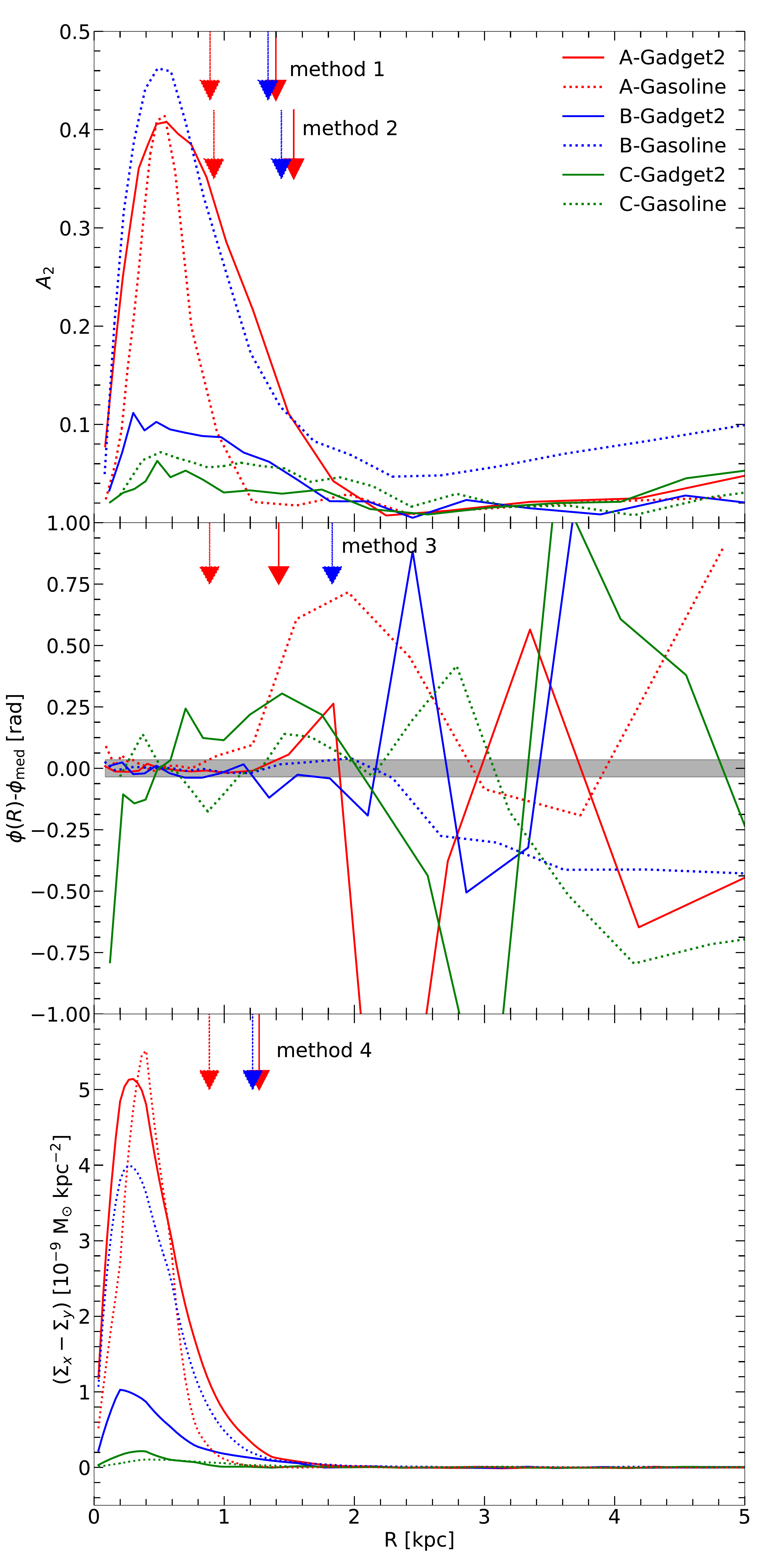}
    \caption{Top: normalised Fourier amplitude of the bi-dimensional mass distribution ($A_{2}$) of the central region of galaxies in function of the cylindrical radius ($R$). Middle: phase of the stellar bi-dimensional mass distribution scaled by a mean phase value $\phi(R)-\phi_{\rm med}$ in function of the cylindrical radius ($R$). $\phi_{\rm med}$ is taken as an average of the bar phase at the radius where the $A_{2}$ profile peaks with the two previous and two following bins values. The shadowed region corresponds to the chosen threshold ($\pm 2\degr$). Bottom: difference of the surface stellar density between semi-major and semi-minor axis of the central mass distribution of the galaxies. If the central region is axisymmetric, the difference is close to zero. The arrows on the top of the panels show the bar length measured with the different methods. CLUES-GADGET2 galaxies are plotted with coloured solid lines and CLUES-GASOLINE galaxies are plotted with coloured dotted lines. We choose red, blue and green for galaxies A, B and C, respectively.}
    \label{fig:barstrength}
\end{figure}

\section{Analysis and results}
\label{sec:analysisresults}

Fig.~\ref{fig:bars} shows, at redshift $z=0$, face-on (left panels) and edge-on (right panels), the projected stellar mass density distribution $\Sigma$ of our six simulated galaxies (see labels) in a box of side $10\,{\rm kpc}$. Images were generated using the package \texttt{Py-SPHViewer} \citep{sphviewer}, coloured according to their projected stellar mass density using a logarithmic scale, and rotated to have the total stellar angular momentum aligned with the z-axis.

Visual inspection of these figures shows that at least three of the studied galaxies present an elongate structure in the centre.

\subsection{Bar properties}
\label{subsec:bars}

Following \citet{Athanassoula_etal2013} we identify bars in our simulated galaxies, computing the amplitudes through the $m=2$ Fourier modes of the face-on projected stellar mass distribution (Fig.~\ref{fig:bars}, left panels), defined as follows:

\begin{equation}
\label{ec:a2_b2}
\begin{split}
        a_{2}(R) & =\frac{1}{M(R)}\sum_{i=1}^{N(R)}M_{i}\:cos(2\phi_{i}) \\
        b_{2}(R) & =\frac{1}{M(R)}\sum_{i=1}^{N(R)}M_{i}\:sin(2\phi_{i}). 
\end{split}
\end{equation}

Here, $R$ is the cylindrical radius of a ring defined by distances in the range [$R,\:R+dR$], $\phi_i$ and $M_i$ are the azimuthal angle and the stellar mass of the i$^{th}$ particle inside the ring, respectively; $N(R)$ is the number of particles within the ring and $M(R)=\sum_{i=0}^{N(R)}M_i$ its corresponding mass. Then, the bar strength can be quantified through the normalised amplitude of the $m=2$ mode:

\begin{equation}
        A_2(R) = \sqrt{a_2^2(R) + b_2^2(R)} 
        \label{ec:A2}
\end{equation}

With this definition, for a perfectly aligned distribution, as in a bar, $A_2=1$ given that $a_2=~\int_{0}^{2\pi}\cos(2\phi) d\phi =\cos(2 {\phi}_{0})$ and $b_2=\int_{0}^{2\pi} \sin(2 \phi) d\phi=\sin(2\phi_0)$, where $\phi_0$ is the bar position angle.

In Fig.~\ref{fig:barstrength}, we show the bar strength parameter $A_{2}(R)$ (top panel) for our six simulated galaxies, computed in 20 cylindrical equal-number bins\footnote{The bins are built in order that each bin contains the same number of particles.} in the range $0<R<6\,{\rm kpc}$. This means between $\sim \! 8,000$ particles per bin for the less massive galaxy, and up to $\sim \! 32,000$ for the most massive one. 

In the top panel of Fig.~\ref{fig:barstrength} we can see three high peaks; galaxies A-Gadget2, A-Gasoline and B-Gasoline, with a radial profile with a maximum of $A_2 > 0.4$ in their central regions, while the remaining three galaxies (B-Gadget2, C-Gadget2 and C-Gasoline) show a relatively flat profile with $A_2 < 0.1$.

Following \citet{Algorry_etal2017}, we classify galaxies as barred or unbarred based on the $A_2(R)$ maximum value $A_2^{\rm max}$; if $0<A_2^{\rm max}<0.2$, the galaxy is classified as unbarred, while if $0.2<A_2^{\rm max}<1.0$, it is a (weakly or strongly) barred galaxy. Thus, A-Gadget2, A-Gasoline and B-Gasoline are barred galaxies and B-Gadget2, C-Gadget2 and C-Gasoline are unbarred, which we confirmed previously by visual inspection of the face-on projections of stellar mass density (Fig.~\ref{fig:bars}(a)).

\subsubsection{Determination of bar length}
\label{subsubsec:barlength}

Bar length $l_{\rm{bar}}$ is one of the most important properties of barred galaxies. There is no single way to define the length of a bar; \citet{Athanassoula&Misiriotis2002} present a compilation of different methods to estimate bar length. \citet{Scannapieco&Athanassoula2012} implemented some of these methods in cosmological numerical simulations, showing general agreement between them. 
In this work, we have used four different methods to estimate bar length: the three methods used in \citet{Scannapieco&Athanassoula2012} plus the method implemented by \citet{Algorry_etal2017}.

\begin{itemize}
    \item[-] \textit{method 1}: following \citet{Algorry_etal2017}, the first method consists in estimating the bar length on the $A_2(R)$ profile where the curve drops below 0.15 after reaching its maximum ($A_2^{\rm max}$).
    
    \item[-] \textit{method 2}: the second method is similar to the first; instead of using a fixed threshold for all galaxies, it uses a fraction of the maximum value ($A_2^{\rm max}$). In theory, if the disc and the bar were rigid bodies, the bar length would be the radius where the $A_2(R)$ profile drops to zero. However, this does not occur in simulated galaxies; therefore, we must choose this value arbitrarily. Following \citet{Scannapieco&Athanassoula2012}, we choose this value as $25\%$ of the $A_2^{\rm max}$.

    \item[-] \textit{method 3}: the third method uses the azimuthal angle radial profile $\phi(R)$ to estimate bar length. The azimuthal angle is calculated from the face-on projected mass distribution (Eq.~\ref{ec:a2_b2}), where $\phi(R) = 0.5 \arctan(b_2/a_2)$. Theoretically, if we consider the bar as a solid body, the position angle of the bi-dimensional mass distribution should be constant along the bar. This is not completely true for simulations or observations, as the phase of the bar varies little along the bar, so the phase can be considered within a certain tolerance threshold. \citet{Athanassoula&Misiriotis2002} and \citet{Scannapieco&Athanassoula2012} use a threshold of $\pm arcsin(0.3)$, which is chosen ad hoc. We chose a smaller threshold  ($\pm arcsin(0.035)$) to improve the quality of the bar length determination.
    As seen in the middle panel of Fig.~\ref{ec:A2}, the first large fluctuations of the bar phase determine the bar size.
    Looking at the curves, at small radii ($R \lesssim 1.5\,\rm{kpc}$) the bar position angle is almost constant with very small fluctuations contained inside the grey shaded area.  At large radii ($R \gtrsim 1.5\,\rm{kpc}$) the fluctuations increase their amplitude considerably. Such large oscillations indicate that the position angle of the bar is not well defined and point out that the bar has ended and the disc is the dominant component.
    
    \item[-] \textit{method 4}: The fourth method estimates the bar length using the difference between stellar density profiles along semi-major and semi-minor axes of the bar. In the centre of the galaxies, the density profile along both axes must coincide and, as we move along the bar, the difference between these profiles must be increasing. Where the bar ends, these profiles must be equal again (if we consider an axisymmetric disc). This theoretical argument is not completely valid in cosmological simulations. The profiles will tend to approximate each other but they never will be equal. Following \citet{Scannapieco&Athanassoula2012}, we take the bar length as the radius where the difference between the profiles drops to $5\%$ of their maximum value.

\begin{table}
\caption[]{Values of bar length at $z=0$ in units of kpc measured with the different methods.}
\label{tab:bar_length}
\centering
\begin{tabular}{lccc}
\hline \noalign{}
  & A-Gadget2 & A-Gasoline & B-Gasoline \\
\hline \noalign{}
\textit{method 1} & 1.40 & 0.89 & 1.34 \\
\textit{method 2} & 1.53 & 0.92 & 1.44 \\
\textit{method 3} & 1.42 & 0.89 & 1.83 \\
\textit{method 4} & 1.27 & 0.89 & 1.22 \\
\hline 
\end{tabular}
\end{table}

The resulting bar length measurements are marked with coloured arrows in Fig.~\ref{fig:barstrength}. The bar length values are also quoted in Table~\ref{tab:bar_length}. 
Table~\ref{tab:properties} lists the mean values of bar length as the average of the four methods. The errors of bar length are the maximum difference between each method to the mean value.
We can see from Table~\ref{tab:bar_length} that all the values are in good agreement.
It should be noted that the method that results in a shorter (or longer) bar length estimation, does not necessarily result in a shorter (or longer) bar length estimation in another galaxy too.
    
\end{itemize}

\subsubsection{Pattern speed estimation}

\begin{figure}
        \includegraphics[width=\columnwidth]{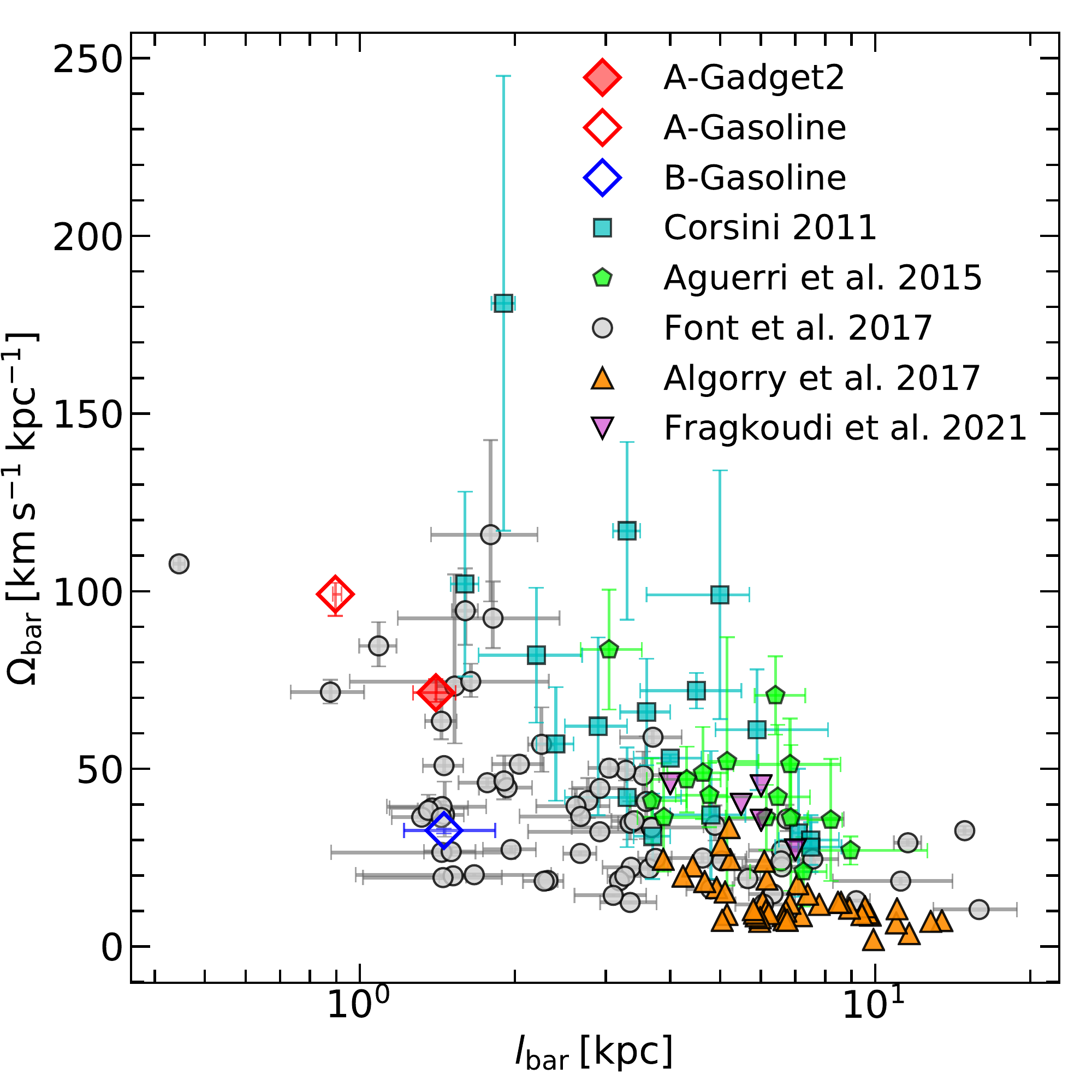}
    \caption{Pattern speed of the barred galaxies $\Omega_{\rm bar}$ as function of the bar length $l_{\rm bar}$. The coloured diamonds show the mean of the four values of $\Omega_{\rm bar}$ and $l_{\rm bar}$ estimations for our simulated galaxies. The error bars show the maximum difference between the mean and the four measurements. Grey circles shows observational data of \citet{Font_etal2017}, green pentagons are galaxies from \citet{Aguerri_etal2015}, cyan squares are galaxies from \citet{Corsini_2011}, orange triangles are galaxies of EAGLE simulations from \citet{Algorry_etal2017}, pink inverted triangles are galaxies of AURIGA simulations from the work of \citet{Fragkoudi_etal2021}.}
    \label{fig:omegalbar}
\end{figure}

Another characteristic property of bars is the pattern speed $\rm{\Omega_{bar}}$. The pattern speed determines how fast the bar rotates. Theoretically, the bar should behave as a solid body, but in practice this is not true. There are many contaminating particles from the disc and the bulge. To avoid this kind of particle, we have taken the pattern speed averaging the angular velocity $\Omega(R)= V_{\phi}/R=(xv_y-yv_x)/(x^2+y^2)$ of star particles at the extremes of the bar: $l_{\rm bar}\pm 0.15\,{\rm kpc}$, $\phi_{\rm med} \pm 5\degr$ and $z \pm 1\,{\rm kpc}$. 
Since we have four values for $l_{\rm bar}$ from the different methods, we calculate $\rm{\Omega_{bar}}$ for each bar length and average them. We take the errors as the maximum absolute difference between the individual pattern speed and the mean value.
The values of $\rm{\Omega_{bar}}$ are shown in Table~\ref{tab:properties}.

In Fig.~\ref{fig:omegalbar}, we plot ${\rm \Omega_{bar}}$ as a function of the bar length $l_{\rm bar}$ for our three barred galaxies (coloured diamonds) and compare them with observational results from \citet{Font_etal2017, Aguerri_etal2015} and \citet{Corsini_2011}. We also compare them to the theoretical predictions of \citet{Algorry_etal2017} and \citet{Fragkoudi_etal2021} for barred galaxies in the EAGLE and AURIGA simulations respectively. This figure shows that our simulated bars occupy the short tail-end of $l_{\rm bar}$ in the observed bar length distribution, contrary to earlier claims that simulations often produce long bars but rarely short ones \citep{Erwin_2005}. They are also systematically shorter by a factor $\sim \! 5$ than the typical values obtained in the EAGLE or AURIGA cosmological simulations \citep{Algorry_etal2017, Fragkoudi_etal2021}. It also shows that the pattern speeds are in good agreement with the range obtained from observational results of \citet{Font_etal2017}, with A-Gadget2 and A-Gasoline galaxies rotating fast and B-Gasoline rotating at a much slower rate.

\begin{figure}
        \includegraphics[width=\columnwidth]{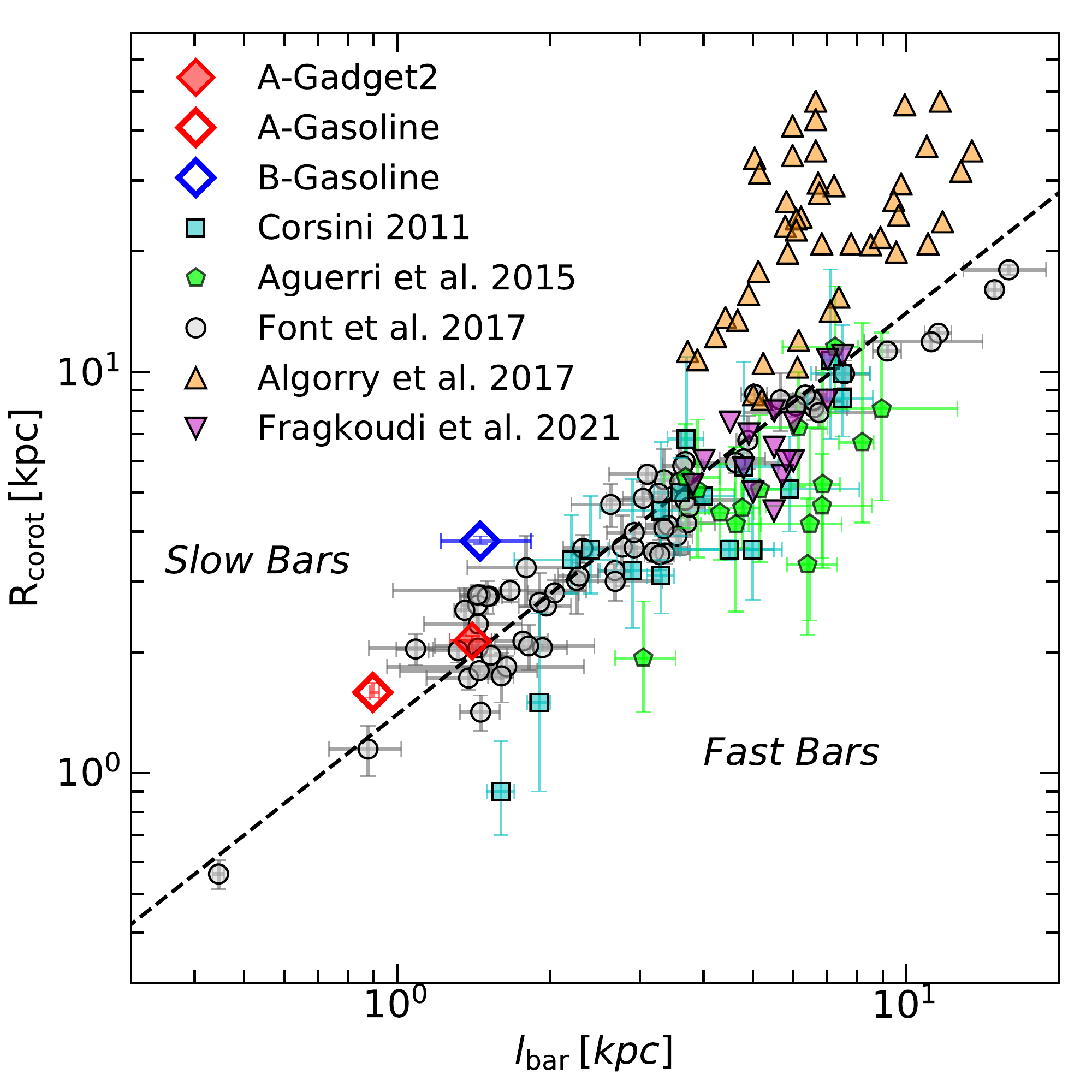}
    \caption{Corotation radius $R_{\rm{corot}}$ as function of bar length $l_{\rm bar}$. The diamonds show the mean of the four values of $R_{\rm corot}$ and $l_{\rm bar}$ estimations and the error bars show the maximum difference between the mean and the four measurements. See labels on the plot.
    The dashed line represents the correlation $R_{\rm{corot}} = 1.4 l_{\rm{bar}}$ that divides the barred galaxies into slow and fast rotators.}
    \label{fig:Rcor_lbar}
\end{figure}

As seen in Fig.~\ref{fig:omegalbar} rotation speeds depend on the bar length and so, to know if a bar is rotating fast or slow according to its length, $\rm{\Omega_{bar}}$ is usually expressed in terms of the quotient of the corotation radius $R_{\rm{corot}}$ and the bar length $l_{\rm bar}$.                                                                                                  
In Fig.~\ref{fig:Rcor_lbar}, we plot the corotation radius $R_{\rm corot}$ as a function of bar length $l_{\rm{bar}}$ using the same symbol and colour coding as in Fig.~\ref{fig:omegalbar}. 
As for $\rm{\Omega_{bar}}$, we calculate four values for the corotation radius and take the average as the $R_{\rm corot}$ for each galaxy. The error bars correspond to the maximum difference between each measurement and the mean.
The dashed line shows the relation $R_{\rm corot}=1.4 l_{\rm bar}$, which is usually implemented to classify bars as \textit{slow} or \textit{fast} rotators \citep{Debattista&Sellwood_2000}. 
In this plot, we can see that our simulated bars fit the short tail-end of fast bars, complementing the results from \citet{Fragkoudi_etal2021} and in good agreement with observational data.

\subsection{Bar temporal evolution}
\label{subsec:tempevol}

\subsubsection{Onset of the bar}

\begin{figure}
        \includegraphics[width=\columnwidth]{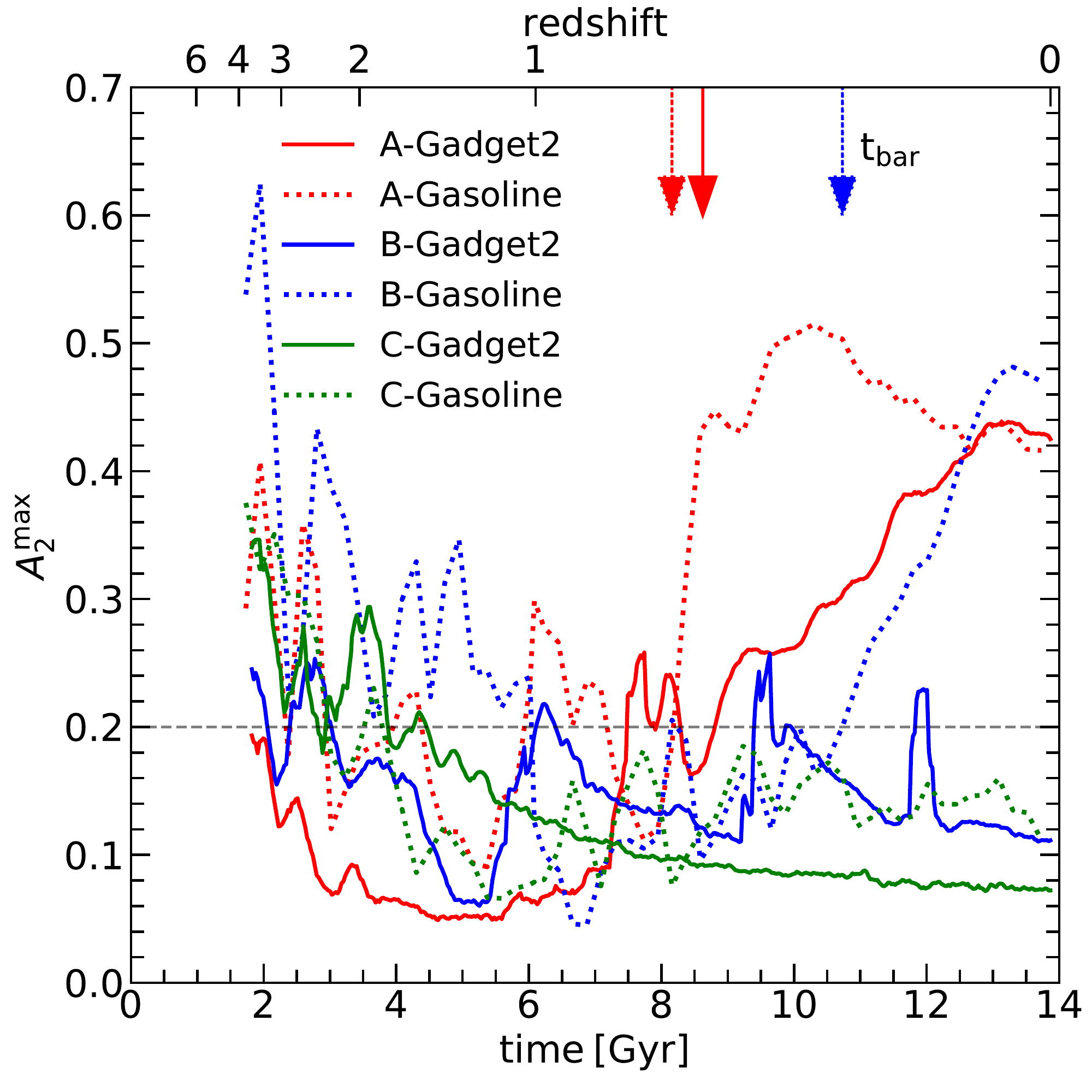}
    \caption{Bar strength $A_{2}^{\rm max}$ as a function of time. We use the same colour coding as Fig.~\ref{fig:barstrength}. We consider that a galaxy has a bar when the $A_{2}^{\rm max}$ parameter exceeds the limit of $A_{2}^{\rm max}=0.2$ (grey dashed horizontal line). The arrows on the top of the plot show the time at which the $A_{2}^{\rm max}$ parameter crosses the grey dashed line for the last time (bottom up); therefore, we take this time as the bar formation time $t_{\rm bar}$. Note that a running average algorithm has been applied to smooth the curves.}
    \label{fig:A2max_time}
\end{figure}

As well as measuring the bar strength at $z=0$ to determine the bar presence, we can analyse its evolution to estimate the bar formation time (see Fig.\ref{fig:A2max_time}).
Although bar formation is a process rather than an event, during a galaxy’s lifetime a characteristic time of bar formation can be defined using these curves. Following \citet{Algorry_etal2017}, we define the bar formation time $t_{\rm bar}$ as the time when the bar strength parameter crosses the threshold $A_2^{\rm max} = 0.2$ for the last time bottom up. In the practice, we select bars at $z=0$, follow these bars back in time and see the last time when the $A_2^{\rm max}$ parameter crossed the 0.2 threshold and select this time as the bar time formation $t_{\rm bar}$. 
In Fig.~\ref{fig:A2max_time}, we show the bar strength parameter in function of time for our simulated galaxies. At early times, $A_2^{\rm max}$ parameter shows a noisy behaviour, because of mergers and the fly bys that the galaxy suffers during its formation. After this period, the strength parameter shows a rising profile for barred galaxies, but remains approximately constant for unbarred ones.

Vertical arrows show $t_{\rm bar}= 8.69$, $8.09$ and $10.62 \,{\rm Gyr}$ for A-Gadget2, A-Gasoline and B-Gasoline respectively, while for the three unbarred galaxies, at late times ($t \gtrsim 8\,{\rm Gyr}$), $A_2^{\rm max}$ are always below the $0.2$ limit.
The three bars develop in the last 3-6~Gyr after the accretion epoch. The bar of A-GASOLINE grows rapidly ($\sim \! 1\,{\rm Gyr}$), while the bars of A-GADGET and B-GASOLINE take longer ($\sim \!4\,{\rm Gyr}$ and $\sim \!3\,{\rm Gyr}$ respectively) to reach their maximum strength. Once the bar strength reaches its maximum, it remains approximately constant up to the present.

\subsubsection{Analysing disc instability}
\label{subsec:discinstability}

Early numerical simulations of self-gravitating cold discs have shown that they become rapidly unstable against bar formation \citep[][among others]{Miller&Prendergast1968, Hockney&Hohl1969, Miller_etal1970, Kalnajs1972, James&Sellwood1978, Combes&Sanders1981, Sparke&Sellwood1987}. 
\citet{Ostriker&Peebles1973} and \citet{Efstathiou1982} implemented dark matter halos to stabilise these discs, proposing that bars form in stellar discs where self-gravity makes the dominant contribution while unbarred discs are those where the dark matter prevails. They also introduced the ratio between the maximum rotational velocity and the individual disc contribution as a parameter to predict the stability of cold discs.

Accordingly, semi-analytical models of galaxy formation \citep[e.g.][]{MoMao&White1998, Cole_etal2000, Bower_etal2006, Lacey_etal2016} usually implement a slightly modified version of this parameter as the ratio between total and disc circular velocities (both at the half-mass radius) as a bar-instability parameter that makes it possible to distinguish between stable and unstable discs:

\begin{equation}
    f_{\rm{disc}} = \dfrac{V_{c}\rm{(r_{50})}}{\sqrt{G{\rm M_{disc}}/\rm{r_{50}}}}.
    \label{ec:fdisc}
\end{equation}

In our work, we use the version used by \citet{Algorry_etal2017} applied in EAGLE simulations. In Eq.~\ref{ec:fdisc}, $ V_{c}{\rm(r_{50})}$, is the total circular velocity of the galaxy at the half stellar-mass radius (${\rm r_{50}}$), $G$ is the gravitational constant and ${\rm M_{disc}}$ is the galaxy stellar mass. We confirm by a double exponential fit that our galaxies are in fact disc galaxies in order to perform this approximation. Systems with $f_{\rm disc} \lesssim 1.1$ are considered to have gravitationally unstable discs, while those with $f_{\rm{disc}} \gtrsim 1.1$ are stable without developing a barred structure.

While the $f_{\rm{disc}}$ parameter seems to be a good predictor of disc instability, there is evidence that this parameter alone is insufficient to predict bar formation \citep{Athanassoula2008}. Numerical simulations have shown that initially \textit{stable} discs ($f_{\rm{disc}} \gtrsim 1.1$) can develop a bar \citep{Athanassoula&Misiriotis2002, Athanassoula2003}, and even if a disc is \textit{bar unstable} ($f_{\rm disc} \lesssim 1.1$), a bar may not develop because of high velocity dispersion of the dark matter halo. To solve this issue, \citet{Algorry_etal2017} proposed combining the $f_{\rm disc}$ parameter with another parameter that takes into account the global importance of the whole system and not only the local gravitational importance of the disc:

\begin{equation}
    f_{\rm dec} = \dfrac{V_{c}\rm{(r_{50})}}{V_{\rm max, halo}}.
\end{equation}

The $f_{\rm dec}$ parameter measures the decline in the curve of total circular velocity, where $ V_{c}{\rm(r_{50})}$ is the total circular velocity of the galaxy at the half stellar-mass radius and $V_{\rm max, halo}$ is the maximum circular velocity of the dark matter halo. Thus a galaxy with $f_{\rm dec}<1$ is a galaxy whose circular velocity increases beyond that of the disc, in such a way that the velocities of the halo particles will be higher than those of the disc, which may prevent the assembling of the particles on the bar delaying its formation or avoiding it completely. Galaxies with $f_{\rm dec}>1$ are those with declining rotation curves, where the rotation velocity of the disc predominates over the velocity of halo particles. Then, we adopt $f_{\rm dec}=1$ as the threshold to divide between favoured ($f_{\rm dec}>1$) and disfavoured ($f_{\rm dec}<1$) galaxies for bar formation.

\begin{figure}
        \includegraphics[width=\columnwidth]{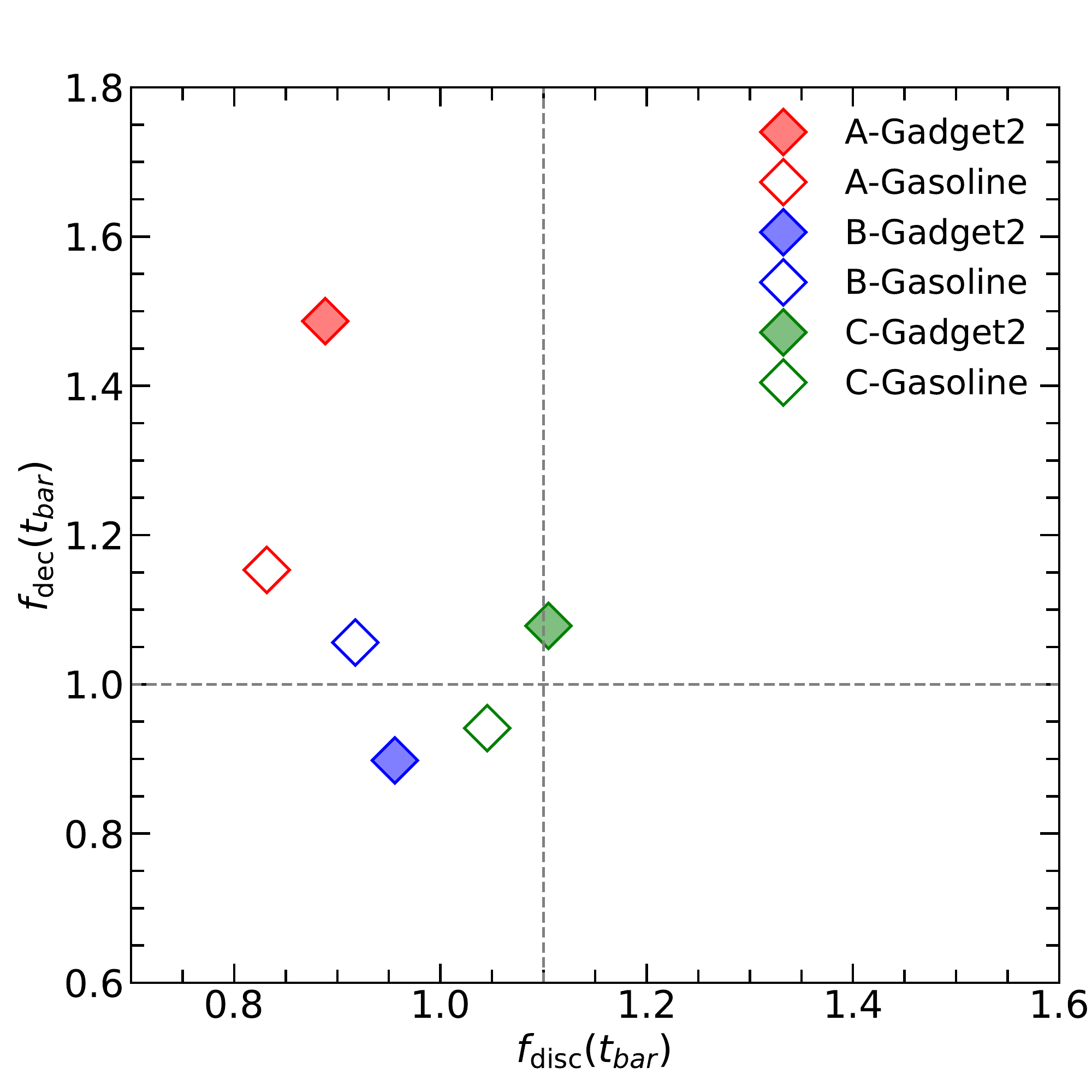}
    \caption{$f_{\rm dec}$ vs $f_{\rm disc}$ parameters. Each parameter is measured at the bar formation time $t_{\rm bar}$. Following \citet{Algorry_etal2017}, for unbarred galaxies we have taken $t=8.8\,{\rm Gyr}$ ($z=0.5$). Horizontal and vertical dashed lines indicate the values $f_{\rm dec}=1.0$ and $f_{\rm disc}=1.1$ thresholds respectively.}
    \label{fig:fdec_fdisc}
\end{figure}

In Fig.~\ref{fig:fdec_fdisc}, we show the $f_{\rm dec}$ vs $f_{\rm disc}$ parameters, both calculated at bar formation time $t_{\rm bar}$. We see that barred galaxies locate in the top left quadrant of the plot, i.e. although each parameter separately is not a good predictor of disc instability, if we combine both parameters, we obtain a good prediction method for \textit{bar unstable} galaxies.

This analysis could explain why even if we have pairs of galaxies obtained from the same initial conditions, one has a bar (B-GASOLINE) and the other does not (B-GADGET2). Similarly, it helps to understand why C-GADGET2, despite having twice the stellar mass of C-GASOLINE, does not develop a bar even though both have similar haloes. 
We expand this discussion at the end of Section~\ref{subsec:formation_time}.

\subsubsection{Bar length and pattern speed evolution}

\begin{figure}
    \includegraphics[width=\columnwidth]{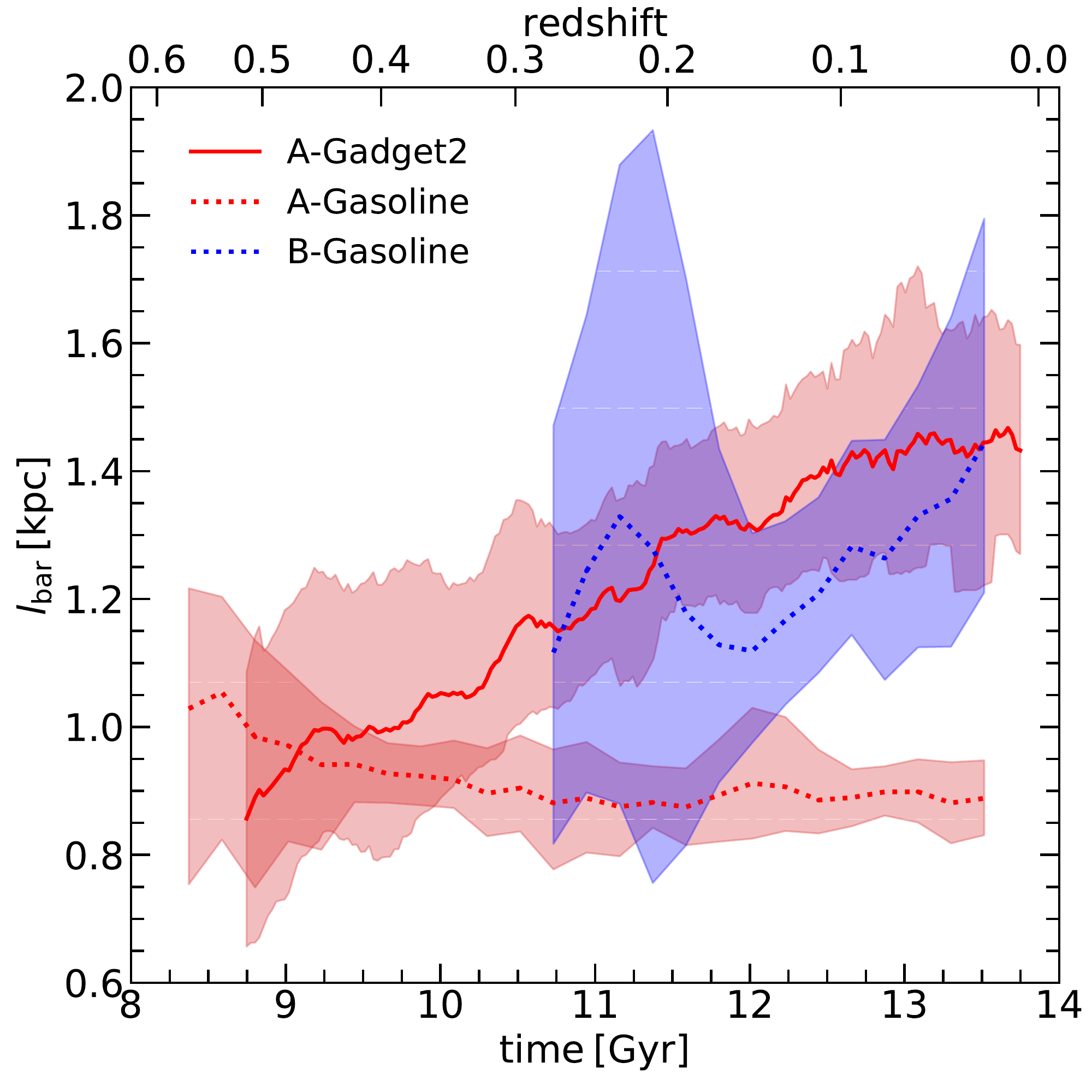}
    \caption{Bar length as a function of time. The curves start at the bar formation time $t_{\rm bar}$} of each galaxy (see Fig.~\ref{fig:A2max_time}). Galaxies A-Gadget2, A-Gasoline, B-Gasoline are plotted with red solid, red dotted and blue dotted lines respectively. Each curve corresponds to the mean $l_{\rm bar}$ estimation. The shaded regions show the maximum difference in bar length between the four methods and the mean. Note that a running average algorithm has been applied to smooth the curves.
    \label{fig:lbar_time}
\end{figure}

\begin{figure}
        \includegraphics[width=\columnwidth]{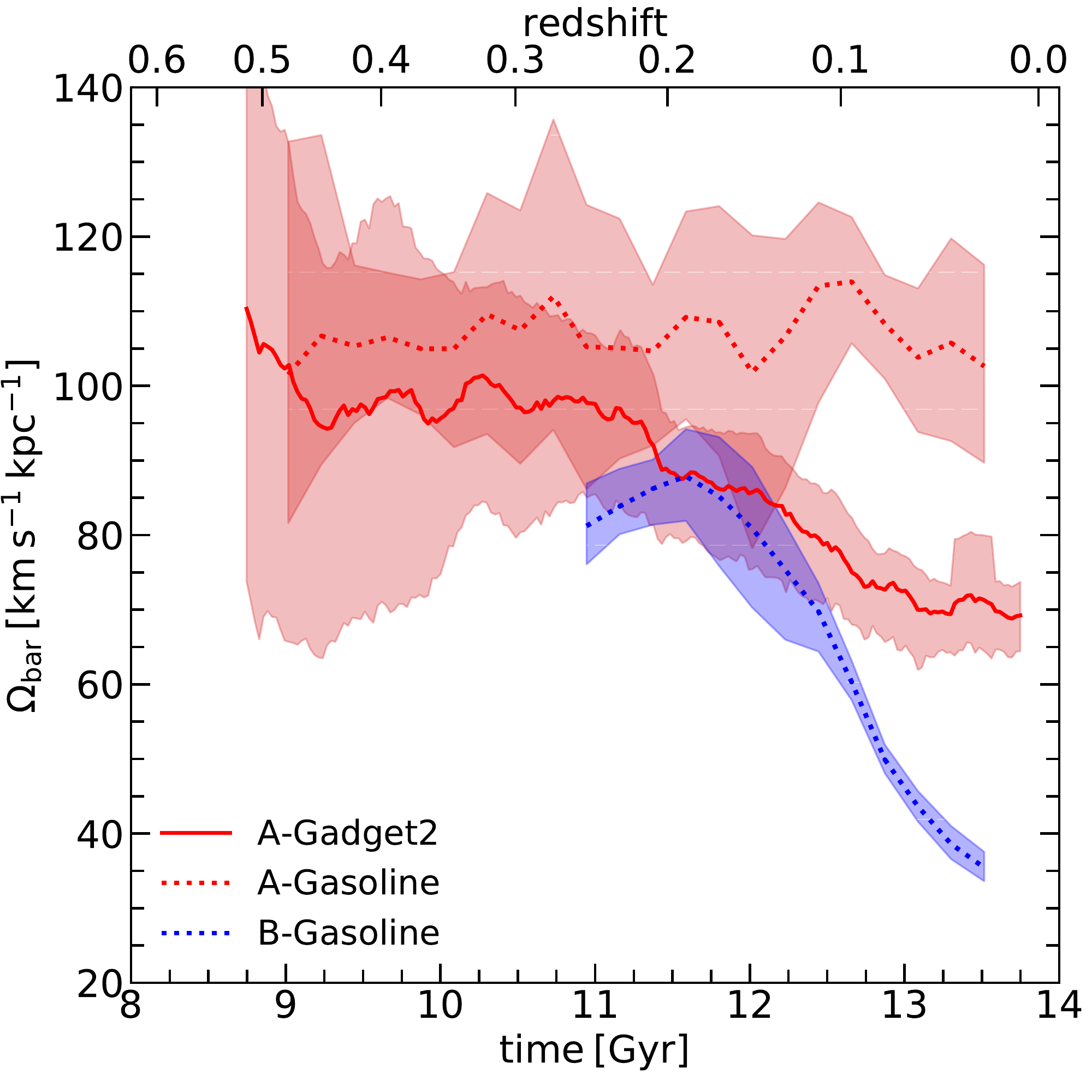}
    \caption{Pattern speed of the bar as a function of time. The curves represent the mean ${\rm \Omega_{bar}}$ of each galaxy between the four estimations of bar length. The shaded region represents the maximum difference between the four measurements of pattern speed for the different estimations of bar length and the mean. Each curve starts at the bar formation time (see Fig.~\ref{fig:A2max_time}). The same colour coding as Fig.~\ref{fig:lbar_time} is used. Here we also applied a running average algorithm to smooth the curves.}
    \label{fig:omega_time}
\end{figure}

Previous works \citep[e.g.][]{Debattista&Sellwood_2000, Athanassoula&Misiriotis2002} have shown that as a bar develops, it grows and slows down.  
We examine this behaviour in Fig.~\ref{fig:lbar_time}, where we plot the temporal evolution of bar length as a function of cosmic time for our three barred galaxies, starting from their formation time ($t_{\rm bar}$) all the way to redshift $z=0$. 
In this figure, we can see that galaxies A-Gadget2 and B-Gasoline increase their length by a factor of 1.3 and 1.7, respectively, while for A-Gasoline, it is fixed almost at a constant value.
Bar growth is accompanied by a consistent slowing down of the pattern speed, as we can see in Fig.~\ref{fig:omega_time}. In this figure, for galaxies A-Gadget2 and B-Gasoline, we see a decreasing pattern speed for a factor of 0.7 and 0.4, respectively, while for A-Gasoline the evolution of ${\rm \Omega_{bar}}$ is nearly flat.

The temporal evolution of the bar length and pattern speed seems to correspond with the evolution of the bar strength ($A_{2}^{\rm max}$, see Fig.~\ref{fig:A2max_time}). After the bar formation time ($t_{\rm bar}$), $A_{2}^{\rm max}$ grows with an approximately constant slope for galaxies A-Gadget2 and B-Gasoline although slightly steeper for B-Gasoline. For A-Gasoline, after $t_{\rm bar}$, the $A_{2}^{\rm max}$ parameter increases rapidly reaching a maximum value and then decreases by only a factor of $\sim\! 20 \% $.  An analogue behaviour is seen in the bar length evolution (Fig.~\ref{fig:lbar_time}) in which both A-Gadget2 and B-Gasoline show a positive slope while A-Gasoline is approximately constant.

\subsection{Formation time of stellar particles forming the bar}
\label{subsec:formation_time}

\begin{figure}
        \includegraphics[width=\columnwidth]{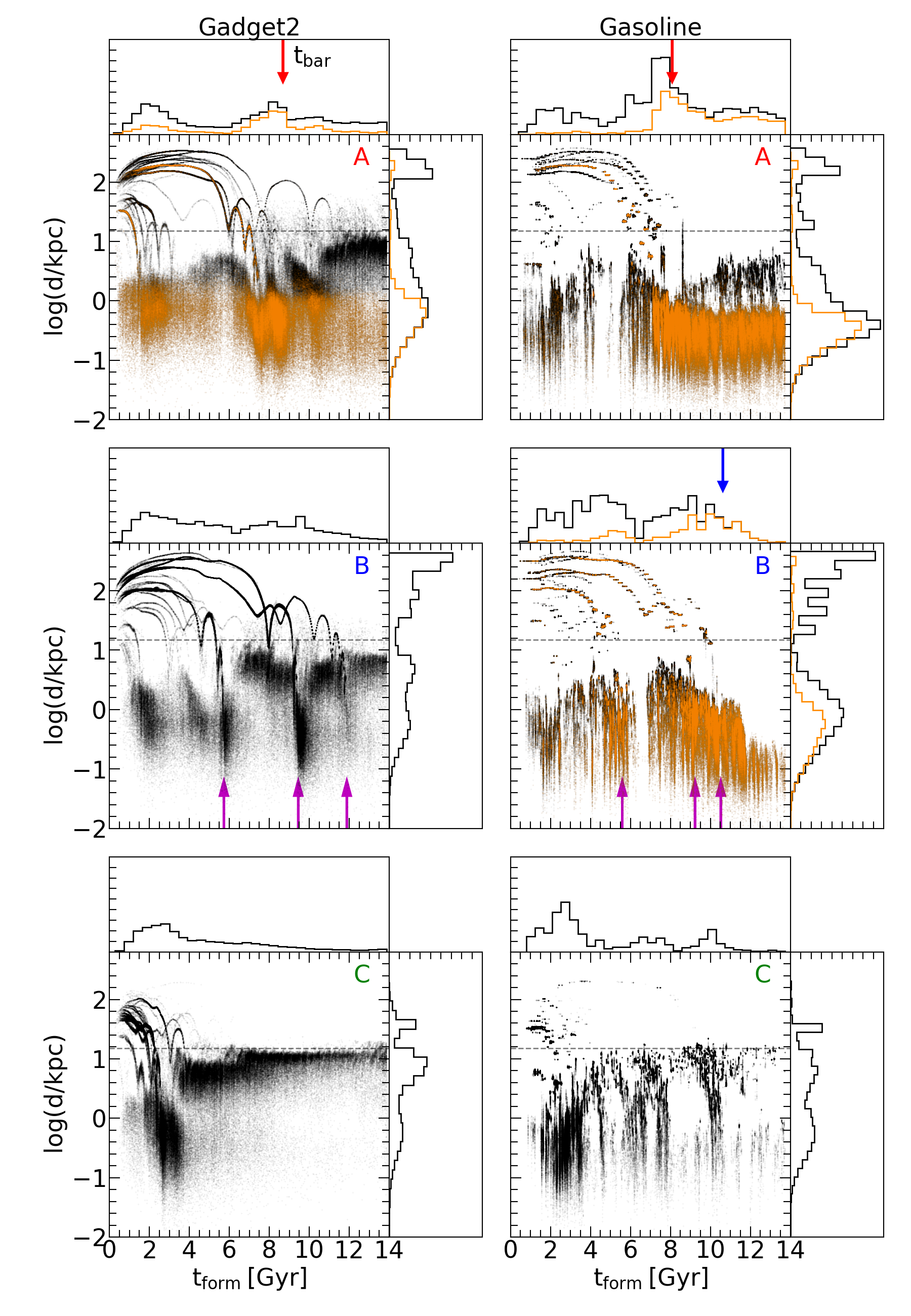}
    \caption{Distance of the stars to the centre of the galaxy at their formation time as a function of that time. The black dots show the total stars inside the galaxy radius and the orange dots show the stars that belong to the bar at $z=0$. The grey dashed horizontal line shows our division between {\it in-situ} ($ d < 15 \, {\rm kpc}$) and {\it accreted} ($d > 15\,{\rm kpc}$) stars. The upper and right histograms are the distribution of both variables. Each tick mark of the histograms correspond to $10^4$ counts, the maximum being 90,000 counts. The vertical arrows at the top indicate the formation time of the bar $t_{\rm bar}$ (see Fig.~\ref{fig:A2max_time} and Table~\ref{tab:properties}), which coincides with the peak of star formation in the galaxy. Vertical magenta arrows in middle panels indicate the time of the merger events of B-galaxies.}
    \label{fig:tform_bar}
\end{figure}

In this section, we analyse the star formation history of the galaxies. In Fig.~\ref{fig:tform_bar} we plot the distance $d$ of star particles from the galaxy centre at their formation time as a function of this time for those stars that belong to the galaxy at redshift $z=0$. 
We classify stars as born {\it in-situ} or {\it accreted}, according to their distance from the centre of the galaxy at the moment of their formation. Distances range between $0 \lesssim d \lesssim 400\,{\rm kpc}$ for all star particles that, at redshift $z=0$, belong to the galaxy $r < {\rm r_{gal}}$ (black dots), or those that belong to the bar (orange dots). 
The horizontal grey dashed line shows our threshold for classifying stars as {\it in-situ} ($ d < 15 \, {\rm kpc}$) or {\it accreted} ($d > 15\,{\rm kpc}$). 
Stars that belong to the bar are defined as those inside a triaxial ellipsoid with semi-axis ratios given by the shape tensor ($T_{ij}=\sum_{k=1}^{N} x_i x_j$ with $i$ and $j$ being the corresponding permutations between x, y and z coordinates and $N$ the total number of particles) computed using all star particles inside the bar length $r < l_{\rm bar}$.
Through this procedure, we find that a significant fraction, about 43\%, 44\% and 35\% of stars belong to the bar for galaxies A-Gadget2, A-Gasoline and B-Gasoline, respectively. These fractions correspond to a stellar mass ${\rm M_{stellar} \sim\! 10^{10}\,{\rm M_{\odot}}}$, which is very close to the mean stellar mass of the \citet{Font_etal2017} sample.

As expected for disc-like galaxies, most stars are born {\it in-situ}, with fractions of 72\%, 47\% and 81\% for galaxies A, B and C of CLUES-GADGET2 simulation respectively and 78\%, 56\% and 86\% for galaxies A, B, C of CLUES-GASOLINE simulation, respectively. Fluctuations in these fractions between the CLUES-GADGET2 and CLUES-GASOLINE runs of each galaxy are $\sim \!10\%$. However, if we focus only on the stars that belong to the stellar bar component at $z=0$, we find that these fractions are very different: 95\%, 96\% and 92\% of stars are born {\it in-situ} for A-Gadget2, A-Gasoline and B-Gasoline, respectively. These fractions show that a very small percentage of the stars that belong at $z=0$ to the bar were formed far away from the main progenitor and accreted later on.

Vertical arrows in the top panels in Fig.~\ref{fig:tform_bar} show bar formation times $t_{\rm bar}$ obtained from Fig.~\ref{fig:A2max_time}. These times seem to correspond approximately with peaks in the star formation times distribution in agreement with previous work from \citet{Fanali_etal2015} pointing out that as the bar forms, the gas reaches the galaxy's most central regions increasing its star formation rate. Once the bar is formed it takes more time for the gas to reach the central region decreasing the star formation rate.
Notice that the $t_{\rm bar}$ also seems to occur near to merger events as seen in Fig.~\ref{fig:tform_bar}. This does not necessarily mean that mergers are the triggers of bar formation but they play an important role on this process. \citet[][]{Moetazedian_etal2017,Zana_etal2017} study the behaviour of bar formation against tidal interactions produced by mergers and flybys. They concluded that close encounters could delay or accelerate the onset of the bar but do not induce the bar formation.

Accretions events could help to understand the different $f_{\rm dec}$ values obtained for B-Gadget2 ($f_{\rm dec}<1$ indicating a rising rotation curve) and B-Gasoline ($f_{\rm dec} > 1$ indicating a declining rotation curve; see Fig.~\ref{fig:fdec_fdisc}). Analysing their circular velocity profiles, at the times we chose to calculate $f_{\rm dec}$, we find that the dark matter halo of B-Gasoline has a slightly declining (nearly flat) profile while B-Gadget2 has a rising one. The rising circular velocity profile of B-Gadget2 seems to be related to its profuse satellite accretion history. 
Indeed, three massive satellites are accreted and merged at 5.73, 9.45, and 11.88~Gyr, respectively (indicated with magenta vertical arrows in Fig.~\ref{fig:tform_bar} middle left panel).
Moreover, although these satellites produce an abrupt increase in the bar strength parameter ($A_{2}^{\rm max} > 0.2$), they are followed by a steady decline to $A_{2}^{\rm max} \sim\! 0.1$, i.e. well below the adopted value to distinguish between barred and unbarred galaxies (see Fig.~\ref{fig:A2max_time} solid blue line where the three peaks are prominent). Notice that although three satellites are also seen for B-Gasoline (see Fig.~\ref{fig:tform_bar} middle right panel) the behaviour of $A_{2}^{\rm max}$ is steadily increasing with time probably due to their different orbit and mass \citep[e.g.][]{Gerin_etal1990,Lokas_etal2016,Peschken&Lokas2019}. Indeed, \citet{Zana_etal2018} argue that the effect of weak interaction could be destructive for the bar formation process.
The main difference between B-galaxies seems to come from the strength of the interaction with the 2nd and 3rd satellites and not from the 1st one.
At the 1st satellite accretion time ($t\sim\! 5.6\,{\rm Gyr}$), the stellar mass difference between B-Gadget2 and B-Gasoline main galaxies, or between both 1st satellites, is only about $10\%$, with this difference increasing up to $40\%$ at redshift $z=0$.
However, the 2nd plus the 3rd satellite have a combined stellar mass of $6.2\times10^9\,{\rm M_{\odot}}$ and  $3.7\times10^9\,{\rm M_{\odot}}$ for B-Gadget2 and B-Gasoline, respectively, which means that their combined mass is about $70\%$ higher for B-Gadget2 than for B-Gasoline.
We consider the combined effect of these two satellites together, given that it is very difficult to disentangle their individual effects on the main galaxy. They accrete approximately at the same time, $t\sim\! 8\,{\rm Gyr}$, and keep interacting, both between them and with the central galaxy, for at least a couple of Gyrs, before merging at different times. Notice that the B-Gasoline bar can finally develop after the accretion of the 3rd satellite; i.e. after the hectic merger activity has somehow stopped or significantly slowed down. The fact that the "2+3" satellite is much more massive in B-Gadget2 than in Gasoline seems to indicate that the bar formation could have been delayed for a much longer period in B-Gadget2 than in B-Gasoline.
Moreover, the last merger event (i.e. the merge of the 3rd satellite) is a much more recent event ($t\sim \!12\,{\rm Gyr}$) for B-Gadget2 while it happened much earlier ($t\sim\!10.5\,{\rm Gyr}$) for B-Gasoline, giving almost twice as much time for bar development.
We would like to stress that many more outputs than those currently available are necessary for CLUES-GASOLINE simulation to enable us to perform a detailed analysis of the tidal interaction effects on these two simulated galaxies.

\section{Summary and conclusions}
\label{sec:summary}

We analyse two zoom-in hydrodynamical numerical simulations of the CLUES project that start at redshift $z=50$ from identical initial conditions and are evolved with {\footnotesize GADGET-2} and {\footnotesize GASOLINE} codes, respectively. The high-resolution region of the simulated volume is an approximately spherical region of $2 h^{-1}\,{\rm Mpc}$ radius with about 53 million equal numbers of gas and dark matter particles. At redshift $z=0$ our simulated galaxies are morphologically classified as discs having from $\sim \!250,000$ to $750,000$ star particles inside (${\rm r_{gal}= 0.15 r_{vir}}$).

Measuring the radial profile of the normalised amplitude of the $m=2$ Fourier mode $A_2(R)$ (Eq.~\ref{ec:A2}), we classify our simulated stellar discs as unbarred ($A_2^{\rm max}<0.2$) or barred ($A_2^{\rm max}>0.2$). We follow the temporal evolution of our galaxies to study the formation and evolution of the stellar bars. At redshift $z=0$, the three most massive haloes of each simulation host a central disc galaxy, with half of them (i.e. three) developing a central rotating stellar bar. In the most massive halo, both simulations (CLUES-GADGET2 and CLUES-GASOLINE) develop a bar; in the second massive halo, only the CLUES-GASOLINE simulation shows a bar, and in the third halo, none of them. 

With total stellar mass in the range $\sim \! 0.5 - 1.4 \times 10^{10}\,{\rm M_{\odot}}$, bar length of $\sim \! 1\,{\rm kpc}$ and pattern speed in the range $32 - 99 {\rm \,km\,s}^{-1}\,{\rm kpc}^{-1}$ these galaxies compare satisfactorily with observational estimates derived for a sample of barred galaxies by \citet{Font_etal2017, Aguerri_etal2015} and \citet{Corsini_2011}. Our bars are short and occupy the short end of the observed bar length distribution, which implies that these high-resolution simulations produce short bars, in contradiction to previous claims that numerical simulations generally produce long bars but infrequently short ones \citep{Erwin_2005}. Our bars are short and not so slow in comparison to previous simulations that raised concerns about the overabundance of slow bars in the $\Lambda$CDM model. Probably, limited numerical resolution plays a pivotal role in angular momentum transfer between bars and dark matter haloes, artificially reducing their pattern speeds. 

We have shown that simulations performed with different codes but starting from identical initial conditions can lead to different bar formation scenarios, because the physics assumed in the simulations can change the final results at small scale where the hydrodynamics of the system prevails over the gravity allowing the bar formation.

In agreement with previous works, bars develop in systems where the disc is gravitationally important compared to the gravitational contribution of the dark matter halo. 
We have confirmed that $f_{\rm disc}$ or $f_{\rm dec}$ parameters alone are not sufficient to predict bar formation, but combining both, we can efficiently predict that the bar formation process will take place.

After their formation, bars grow in length by a factor between 1-2, while their rotation speeds slow down accordingly. Nevertheless, they are not slow when compared to observed pattern speeds, reducing previous concerns about numerical simulations producing systematically slow bars \citep{Algorry_etal2017}.

Also, we see from the last section that most of the stars that belong to the bar at $z=0$ are born inside the galaxy and the formation of the bar causes the triggering of star formation, as found in previous works.

\section*{Acknowledgements}
We are grateful to the anonymous referee for the very constructive and insightful report that has contributed to substantially improving our paper.
We thank the Consejo de Investigaciones Cient\'ificas y T\'ecnicas de la Rep\'ublica Argentina (CONICET) and the Secretar\'ia de Ciencia y T\'ecnica de la Universidad Nacional de C\'ordoba (SeCyT) for supporting this project. 
Our collaboration has been supported by the DFG grant GO 563/24-1 and CONICET. GY acknowledges financial support from the MICIU/FEDER (Spain) under project grant PGC2018-094975-C21. We want to thank Chris Brook for the CLUES-GASOLINE simulation we used in this work. The simulations were performed in the SuperMUC supercomputer at the  Leibniz Rechenzentrum Munich (LRZ) and the MareNostrum supercomputer at the  Barcelona Supercomputing Center (BSC). 
OM thanks Valeria Ricca for her corrections and suggestions that improved the English grammar.

\section*{Data Availability}
The data underlying this article will be shared on reasonable request to the corresponding author.



\bibliographystyle{mnras}
\input{Marioni_etal.bbl}







\bsp        
\label{lastpage}
\end{document}